\documentclass[final, a4paper]{aastex631}

\usepackage{color}
\usepackage{enumitem}
\usepackage{hyperref}
\usepackage{verbatim}
\usepackage{amsmath,amstext,mathrsfs}
\usepackage[all]{hypcap} 
\usepackage{afterpage}    
\usepackage{marginnote}
\usepackage{makecell}
\usepackage{subfigure}
\usepackage{multirow}

\shorttitle{$B-\rho$ slope and ${\cal M}_{\rm A}$}
    \shortauthors{Zhao, Li $\&$ Qiu}
\usepackage{enumitem}
\setlist[enumerate]{listparindent=\parindent}
\makeatletter

\newcommand{\Rmnum}[1]{\expandafter\slowromancap\romannumeral #1@}
\makeatother
\newcommand*{\mk}{}
\newcommand*{\en}{}

\begin{document}

\title{ Slope of Magnetic Field-Density Relation as An Indicator of Magnetic Dominance}

\correspondingauthor{Guang-Xing Li}
\email{gxli@ynu.edu.cn,ligx.ngc7293@gmail.com}

\author[0000-0003-0596-6608]{Mengke Zhao}
\affil{School of Astronomy and Space Science, Nanjing University, 163 Xianlin Avenue, Nanjing 210023, Jiangsu, People’s Republic of China}
\affil{South-Western Institute for Astronomy Research, Yunnan University, Kunming 650091, People’s Republic of China}
\affil{Key Laboratory of Modern Astronomy and Astrophysics (Nanjing University), Ministry of Education, Nanjing 210023, Jiangsu, People’s Republic of China}

\author[0000-0003-3144-1952]{Guang-Xing Li}
\affil{South-Western Institute for Astronomy Research, Yunnan University, Kunming 650091, People’s Republic of China}

\author[0000-0002-5093-5088]{Keping Qiu}
\affil{School of Astronomy and Space Science, Nanjing University, 163 Xianlin Avenue, Nanjing 210023, Jiangsu, People’s Republic of China}
\affil{Key Laboratory of Modern Astronomy and Astrophysics (Nanjing University), Ministry of Education, Nanjing 210023, Jiangsu, People’s Republic of China}

\begin{abstract}
The electromagnetic field is a fundamental force in nature that regulates the formation of stars in the universe. 
Despite decades of efforts, a reliable assessment of the importance of the magnetic fields in star formation relations remains missing. 
In star-formation research, our acknowledgment of the importance of magnetic field is best summarized by the  
\citealt{2010ApJ...725..466C}  $B$-$\rho$ relation 
\begin{equation}
    {\rm log} B(\rho) /{\rm Gauss} = \left\{
    \begin{aligned}
    -5, \,\, {\rm if}\, \rho \, \lesssim \, 10^{-20} \,{\rm g\,cm}^{-3} \\
    \frac{2}{3}\cdot {\rm log} \rho + {\rm log}\rho_0, \,\,  {\rm if} \, \rho \, \gtrsim \, 10^{-20} \,{\rm g\,cm}^{-3} \nonumber
    \end{aligned}
    \right
    .
\end{equation}\
whose interpretation remains controversial \citep{2019ApJ...871...98Z,2021Galax...9...41L,2023ApJ...946L..46C}. The relation is either interpreted as proof of the importance of a magnetic field in the collapse \cite{2019ApJ...871...98Z}, or the result of self-similar collapse where the role of the magnetic is secondary to gravity \cite{2018MNRAS.474.2167L}. 
Using simulations, we find a fundamental relation, ${\cal M}_{\rm A}$-k$_{B-\rho}$(slope of $B-\rho$ relation) relation:
\begin{equation}
    \rm \frac{{\cal M}_{\rm A}}{{\cal M}_{\rm A,c}} = k_{B-\rho}^{\cal K} \approx \frac{{\cal M}_{\rm A}}{7.5} \approx   k_{B-\rho}^{1.7\pm 0.15}. \nonumber
\end{equation}
This fundamental B-$\rho$-slope relation enables one to measure the Alfvénic Mach number, a direct indicator of the importance of the magnetic field, using the distribution of data in the B-$\rho$ plane. It allows us to drive the following empirical  $B-\rho$ relation
\begin{equation} 
    \frac{B}{B_c} = {\rm exp}\left(\left(\frac{\gamma}{{\cal K}}\right)^{-1}\left( \frac{\rho}{\rho_c}\right)^\frac{\gamma}{{\cal K}}\right) \approx \frac{B}{10^{-6.3} {\rm G}} \approx {\rm exp}\left(9 \left(\frac{\rho}{10^{-16.1} {\rm g\,cm^{-3}}}\right)^{0.11} \right)\; \nonumber,
\end{equation}
which offers an excellent fit to the Cruther et al. data, where we assume ${\cal M}_{\rm A}-\rho$ relation ($\frac{{\cal M}_{\rm A}}{{\cal M}_{\rm A,c}} = \left(\frac{\rho}{\rho_c}\right)^\gamma \approx {\cal M}_{\rm A}/{7.5} \approx \left(\rho/{10^{-16.1} {\rm g\,cm^{-3}}}\right)^{\rm 0.19}$). 
The foundational ${\cal M}_{\rm A}-{\rm k}_{B-\rho}$ relation provides an independent way to measure the importance of magnetic field against the kinematic motion using multiple magnetic field measurements. Our approach offers a new interpretation of the \citep{2010ApJ...725..466C}, where a gradual decrease in the importance of B at higher densities is implied.


\end{abstract}

\section{Background}

The magnetic field is a fundamental force that plays a crucial role in the evolution of interstellar gas and star formation \citep{2012ApJ...750...13C,2021Galax...9...41L,2023ASPC..534..193P} through magnetic force and it also changes the initial condition \citep{2023NatAs...7..351Y} by affecting the transport of cosmic rays.

Despite decades of research, constraints on the importance of magnetic fields remain sparse. There are two approaches. 
The first one involves the measurement of dimension numbers which reflects the importance of the magnetic fields. One example is the mass-to-flux ratio, which is the ratio between mass and magnetic flux \citep{1978PASJ...30..681N,2004ApJ...600..279C}. 
The second approach is to derive scaling relations between the magnetic field and density\citep{2010ApJ...725..466C,2019ApJ...871...98Z,2023ApJ...946L..46C}, which serve as bridges between theory and observations.  


The empirical magnetic field density relation ($B-\rho$ relation) is an empirical relation describing the magnetic field strength in the interstellar constructed using a sample of regions with Zeeman observations \citep{2010ApJ...725..466C,2019ApJ...871...98Z,2020ApJ...890..153J,2023ApJ...946L..46C}. The relation can be described using different functions, such as
\begin{equation}
    {\rm log} B(\rho) /G = \left\{
    \begin{aligned}
    -5, \,\, {\rm if}\, \rho \, \lesssim \, 10^{-20} \,{\rm g\,cm}^{-3} \\
    \frac{2}{3}\cdot {\rm log} \rho + {\rm log}\rho_0, \,\,  {\rm if} \, \rho \, \gtrsim \, 10^{-20} \,{\rm g\,cm}^{-3} 
    \end{aligned}
    \right
    .
\end{equation}
 The classical $B-\rho$ relation is an empirical relation, which does not explain how the B-field evolves with density growing. 
\citealt{2019ApJ...871...98Z,2023ApJ...946L..46C} explain a part of classical $B-\rho$ relation with the $\frac{2}{3}$ power-law exponent being caused by the collapse of star formation 
This exponent of $B-\rho$ in star-forming can be linked to the density profile of the region, where $B\propto \rho^{1/2}$ is related to the density profile of $\rho\sim r^{-2}$ \citep{2018MNRAS.474.2167L}.
The $B-\rho$ relation is affected transiently by galactic magnetic field morphology \citep{2024arXiv240210268K}.

{\mk
The magnetic field-density relation, although widely cited, is hard to interpret because both the magnetic field and gravitational force are long-range forces where the scale information is critical \citep{2018MNRAS.474.2167L}. For example,  the energy balance of a cloud reads $B^2r^3 \, \approx \rho^2 r^5$, where the scale information is an essential part. Without understanding the full relationship between $B$, $\rho$, and $r$, due to the missing scale information, constraining the importance of the magnetic field using the \citet{2010ApJ...725..466C} relation is hard. 

One solution, as \citet{2018MNRAS.474.2167L} have pointed out, is to perform joint analysis by combining the magnetic field-density relation and the density-scale relation (density profile). However, this analysis is hard to perform due to the requirement of different types of observations measured on similar scales.  The approach we take in this paper is to determine the importance of the magnetic field by linking density variations with the variations in the strength of the magnetic field. We attempt to link the slope of the $B-\rho$ relation to the importance of the magnetic field measured in terms of the Alfvén Mach number ${\cal M}_{\rm A}$. This equation reflects a fundamental property of the MHD turbulence. This approach also allows for re-interpreting the \citep{2010ApJ...725..466C} relation, as a combination of a fundamental $\cal{M}_{A}-B-\rho$ slope relation, which reflects a fundamental property of compressive MHD turbulence, and the $\rho$-$\cal{M}_A$ relation, which reflects the density dependance of the importance of the magnetic field. 
}

\section{Alfvénic Mach Number \& Fundamental $B-\rho$ Relation}

\subsection{Slope of $B-\rho$ Relation: An Indicator of ${\cal M}_{\rm A}$ }\label{sect2.1}

\begin{figure}
    \centering
    \includegraphics[height= 6.8cm]{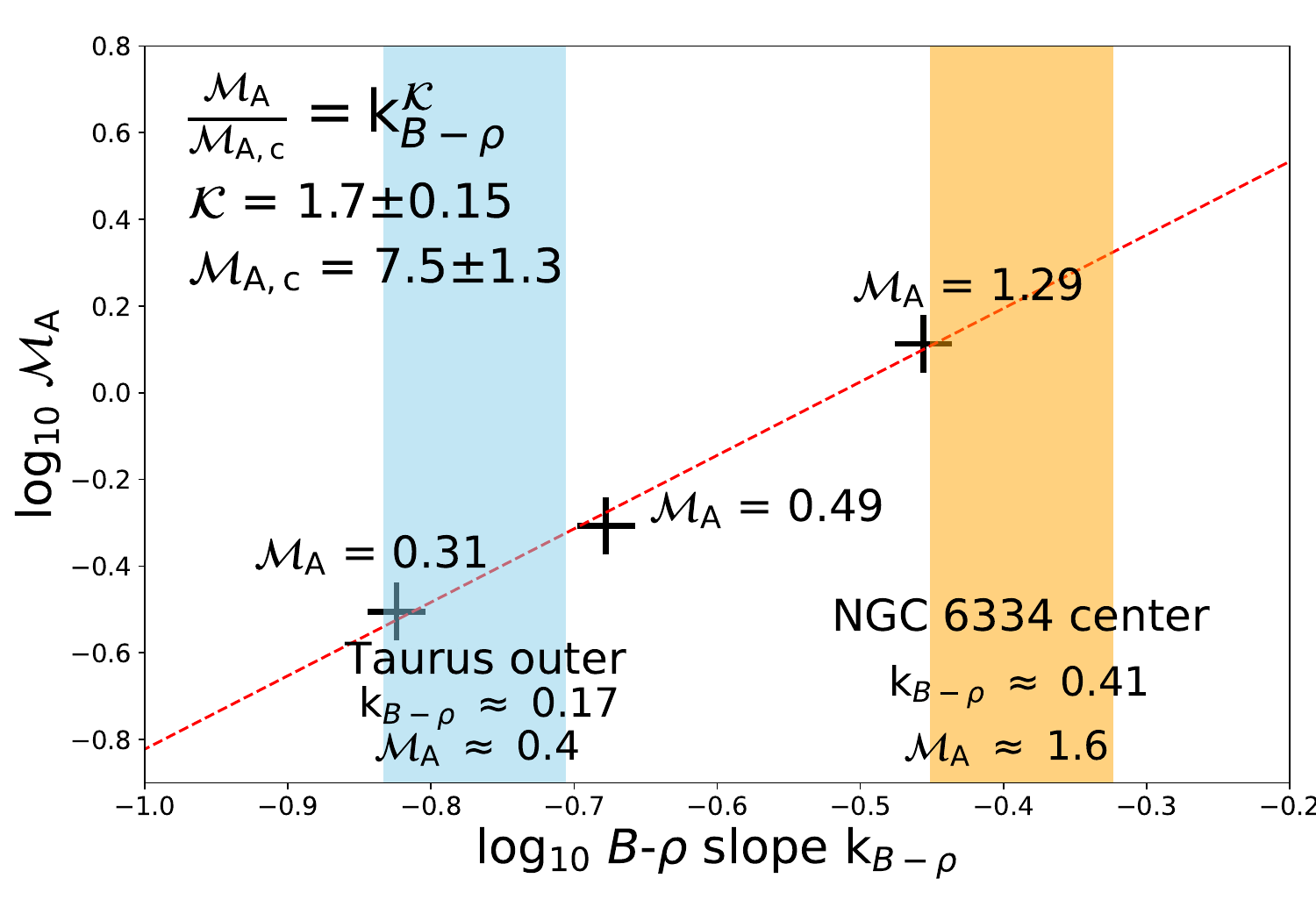}
    \includegraphics[height = 7cm]{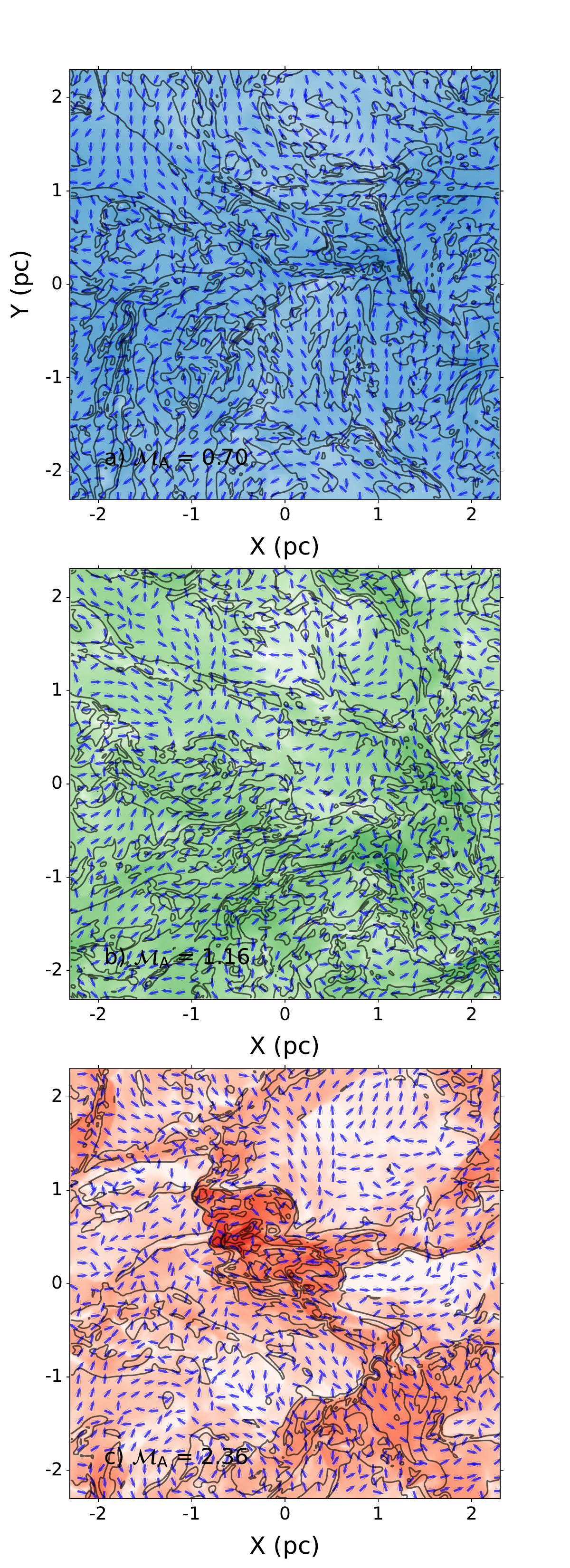}
    \caption{\textbf{The B-$\rho$ slope k$_{B-\rho}$ and Alfvén Mach number ${\cal M}_{\rm A}$ relation.}
    The crosses in the left pane; show the position of three re-scaled simulations with $\beta_0$ as 0.2, 2, and 20, in k$_{B-\rho}$ and ${\cal M}_{\rm A}$ plane, which the k$_{B-\rho}$ is the B-$\rho$ slope of the mean density in various simulations.
    The red dash line shows fitted B-$\rho$ slope k$_{B-\rho}$ and Alfvén Mach number ${\cal M}_{\rm A}$ relation.
    The blue and orange columns show the k$_{B-\rho}$ position of the previous observation, Taurus and NGC 6334 center region, respectively.
    The right panels depict the spatial distribution of the magnetic field (B-field) in a two-dimensional x-y slice, which comes from simulations with different $\beta_0$ and Alfvén Mach number ${\cal M}_{\rm A}$. 
    The color-coded maps represent these simulations as follows: the blue map corresponds to $\beta = 0.2$ and Alfvén Mach number ${\cal M}_{\rm A}$ = 0.31, the green map to $\beta = 2$ with ${\cal M}_{\rm A}$ = 0.49, and the red map to $\beta_0$ = 20 with ${\cal M}_{\rm A}$ = 2.36.
    The vector arrows illustrate the local magnetic field orientations.
    The background contours illustrate the density distribution, ranging from 10$^{-21}$ to 10$^{-18}$ g\,cm$^{-3}$.
    }
    \label{figMAslope}
\end{figure}


To establish ration between $B-\rho$ and ${\cal M}_{\rm A}$, we select a numerical magnetohydrodynamic (MHD) simulation \citep{2012ApJ...750...13C,2015ApJ...808...48B}. 
The initial conditions are generated by a PPML code without self-gravity  \citep{2009JCoPh.228.7614U},  after which the main simulation was performed using AMR code Enzo\citep{1995CoPhC..89..149B,2004astro.ph..3044O} where self-gravity is included. 
In both simulations, the equation of state is isothermal. 
{\mk These simulations are performed with different degree of magnetization (strong $\beta_0$ = 0.2, medium $\beta_0$ = 2 and weak $\beta_0$ = 20) and similar velocity fluctuations ($v_{\rm rms}$ = 9 $c_s$).}

The relatively short simulation time ($0.8\,\rm Myr$) means the effect of self-gravity is only noticeable in the regions of the highest density 
\footnote{The simulations were terminated at around  0.76 Myr, The density after which self-gravity can have an effect should be around 2.6$\times$10$^{-20}$ g\,cm$^{-3}$ (estimated using $\rho_* = 1/(t_*^2\cdot G$), where $t_*$ is the time at which the simulation terminates), which is far above the mean density (3.8$\times$10$^{-21}$) of simulations and this density. {\mk When gravity dominates, it can drive turbulence and increase $\cal{M}_A$, however, the relationship between $B-\rho $ slope and  $\cal{M}_A$ remains unchanged. We have verified this using simulations taken at different times. }}. 
The observed $B-\rho$  relation mostly reflects the interplay between the magnetic field and turbulence cascade, which is the main focus of this work.

We choose three simulations of different initial degrees of magnetization $\beta_0$ = ${8\pi c_s^2 \rho_0}/{B_0^2}$ = 0.2, 2, 20.  The slopes of the  $B-\rho$ relation can be derived after some averages (Method \ref{enzo}).
The Alfvén Mach number ${\cal M}_{\rm A}$ in various simulations can be calculated by the total energy ratio between the magnetic energy $E_B$ and kinetic energy $E_k$:
\begin{equation}
    {\cal M}_{\rm A} = \frac{\sqrt{4\pi\rho} \sigma_v}{B} =\sqrt{\frac{0.5\rho \sigma_v^2}{B^2/8\pi}} =  \sqrt{\frac{E_{\rm k}}{E_B}} ,
\end{equation}
where the ${\cal M}_{\rm A}$ in strong, medium, and weak magnetic field states can be estimated as 0.31, 0.49, and 1.29, respectively.
The  Alfvénic Mach numbers ${\cal M}_{\rm A}$ are measured at the time of the snapshots (Method \ref{enzo}), which is different from the Alfvénic Mach numbers inferred from the initial conditions, due to the dissipation of turbulence and the amplification of the magnetic field.
{\en The changing slope of $B-\rho$ relation reported here is consistent with the density-dependance of this slope as found in \citep{2024MNRAS.530.3431V}, the variations of the slopes of the   $B-\rho$ relation can be explained as the local variations of  ${\cal M}_{\rm A}$.}

We then plot the Mach number against the measured slope of the $B-\rho$  (see Fig.\,\ref{figslopesims}), to reveal a surprisingly simple relation between the slope of the $B-\rho$ relation (k$_{B-\rho}$ = d (log$_{10}$ $B$) / d (log$_{10}$ $\rho$)) and the Alfvén Mach number ${\cal M}_{\rm A}$,
\begin{equation}\label{eqmaslope}
    \rm \frac{{\cal M}_{\rm A}}{{\cal M}_{\rm A,c}} = k_{B-\rho}^{\cal K} ,
\end{equation}
where the ${\cal K}$ = 1.7$\pm$0.15, and the ${\cal M}_{\rm A,c}$ is the characteristic ${\cal M}_{\rm A}$ ($\sim$ 7.5$\pm$1.3, see Fig.\,\ref{figMAslope}). 
The degree of magnetization quantified using ${\cal M}_{\rm A}$ determines the power-law exponent between density and magnetic field.
The ${\cal M}_{\rm A}-{\rm k}_{B-\rho}$ relation could be a fundamental relation in natural.
An intuitive explanation for the correlation between the stronger magnetic field and shallower magnetic field-density relation is that a strong field can regulate the way collapse occurs, such that the matter accumulates along the field lines, weakening the correlation between B and $\rho$.

\begin{figure}
    \centering
    \includegraphics[width = 13cm]{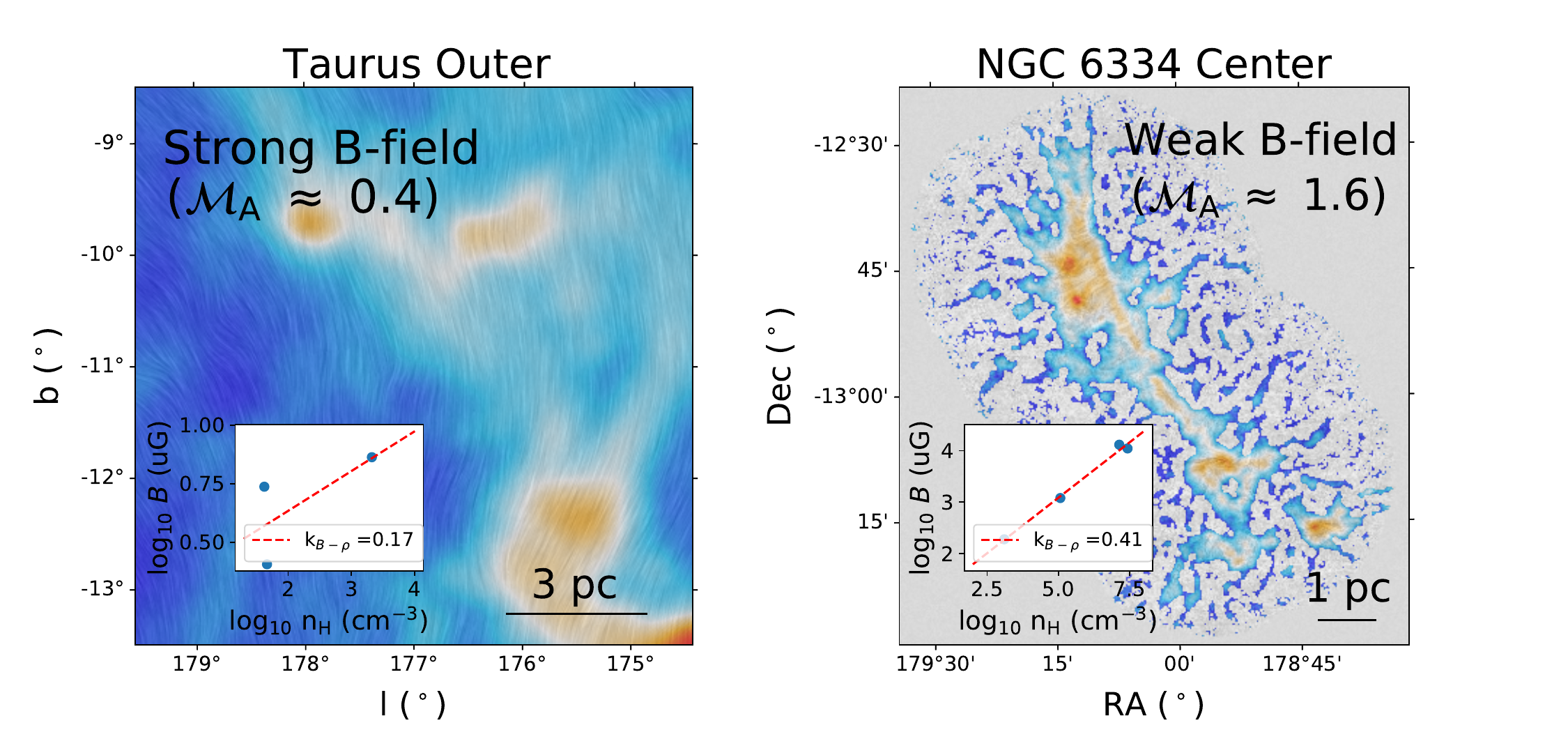}
    \caption{\textbf{B-field and slope of $B-\rho$ relation in Taurus outer and NGC\,6334 center regions.}
    The panels show the magnetic field morphology and slope of $B-\rho$ relation in Tauras outer region and NGC\,6334 center region.
    The $B-\rho$ distribution is derived from Zeeman observations in Tauras outer region \citep{2010ApJ...725..466C,2022Natur.601...49C} and that of NGC 6334 from \citealt{2015Natur.520..518L}.
    The magnetic field morphology is derived by Planck 353 GHz dust polarization \citep{2020A&A...641A..12P} and JCMT 850 $\mu$m dust polarization \citep{2021A&A...647A..78A}.
    }
    \label{figtworegions}
\end{figure}

\subsection{Determining the Importance of B-field using the ${\cal M}_{\rm A}$- k$_{B-\rho}$ relation}


${\cal M}_{\rm A}$ is a fundamental physical quantity that directly characterizes the importance of the magnetic field. 
The fact that this quantity is directly related to the slope of the $B-\rho$ relation provides a new way to study the importance of the magnetic field (see Sect.\,\ref{sect2.1}). 
However, to construct this \emph{single-region} $B-\rho$ relations, we require multi-scales, multi-densities observations towards single regions (Methods\,\ref{apslope}). Another case is the outer region of the Taurus molecular cloud, where multiple constraints of the magnetic field strength using Zeeman splitting \cite{2010ApJ...725..466C,2022Natur.601...49C} are available.


We reveal vastly different roles of the magnetic field (see Fig.\,\ref{figtworegions}) by applying our results to both clouds. 
Toward the Taurus outer region, using Zeeman observations \citep{2010ApJ...725..466C,2022Natur.601...49C}, we estimate a ${\cal M}_{\rm A}$ of 0.8, indicative of a relatively strong (dynamically significant) magnetic field. Towards the NGC\,6334 center region, the ${\cal M}_{\rm A}$ is estimated as 3.0, implying a weak (dynamically weak) magnetic field.

Our assessment of the importance of the magnetic field is consistent with the behavior of these regions as reported in the literature \citep{2022Natur.601...49C,2015Natur.520..518L}: 
The strong magnetic field observed towards the outer region of the Taurus molecular cloud, where  ${\cal M}_{\rm A}$ ($\sim$ 0.4) indicates that the B-field is strong enough to dominate the gas evolution. 
This is consistent with conclusions from previous studies \cite{2016MNRAS.461.3918H,2016MNRAS.462.3602T}. 
At the Taurus outer region, people have observed striations which are taken as signs of the dominance MHD wave \citep{2016MNRAS.461.3918H,2016MNRAS.462.3602T} -- a phenomenon that occurs only if the B-field is strong enough. Towards the dense part where our observations target, it is found that denser gas has lost a significant amount of magnetic flux as compared to the diffuse envelope \citep{2022Natur.601...49C}, and this critical transition, and agrees well with our estimate where $\sqrt{E_B / E_k} \propto {\cal M}_{\rm A}$. 
Towards the central region NGC\,6334, we find that kinetic energy is far above the magnetic energy (${\cal M}_{\rm A} \approx $1.6), dynamically insignificant magnetic field). 
This is consistent with NGC\,6334 being one the most active star-forming regions with a high fraction of dense gas \cite{2022A&A...666A.165S}. 

Using the fundamental ${\cal M}_{\rm A}-{\rm k}_{B-\rho}$ relation (Eq. \ref{eqmaslope}), we can use the slope of $B-\rho$ relation to estimate the Alfvén Mach number ${\cal M}_{\rm A}$, which determine the importance of B-field.




\begin{figure}
    \centering
    \includegraphics[width = 14cm]{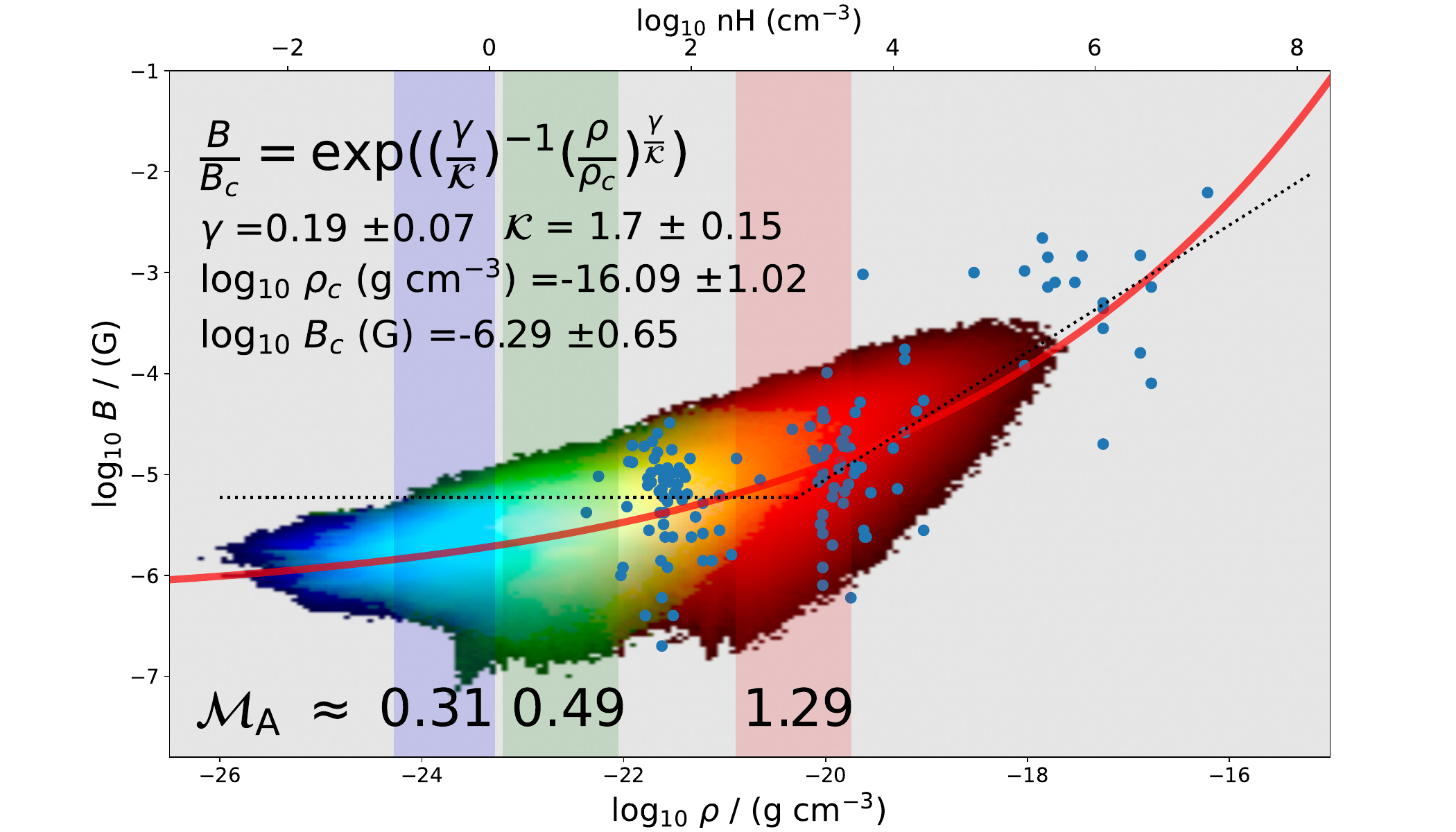}
    \caption{\textbf{$B-\rho$ relation in multi-regions.}
    This panel displays the relation between magnetic field (B) and density ($\rho$) following the exponential function. 
    The red line presents the exponential $B-\rho$ relation derived by ${\cal M}_{\rm A}$-k$_{B-\rho}$ relation and ${\cal M}_{\rm A}-\rho$ relation.
    Within this panel, the blue scatter points represent the distribution of sources measured with the Zeeman effect in the B-$\rho$ space, as detailed in \citep{2010ApJ...725..466C}. 
    The black dashed line shows the old version (power law) $B-\rho$ relation fitted by the Zeeman source (blue scatters).
    The blue, green, and red clouds display the distribution of rescaling simulations with various ${\cal M}_{\rm A}$ as 0.31, 0.49, and 1.29, respectively.
    The blue, green, and red columns show the ${\cal M}_{\rm A}$ and mean density of rescaling simulations with various ${\cal M}_{\rm A}$ as 0.31, 0.49, and 1.29, respectively.}
    \label{figBrho}
\end{figure}

\subsection{Deriving $B-\rho$ relation}


In the past, a convenient way to study magnetic field was to plot the field strength against the gas density, forming the magnetic field-density relation \citep{2010ApJ...725..466C,2019ApJ...871...98Z,2023ApJ...946L..46C}. 
This {\it multi-region} magnetic field-density relation is constructed from observations of different regions which have different sizes and geometries. 
The plural nature of this multi-region magnetic field-density relation makes it not particularly useful in estimating the ${\cal M}_{\rm A}$. 
However, we can decipher this relation using a minimum of assumptions.

We start from the fundamental ${\cal M}_{\rm A}-{\rm k}_{B-\rho}$ relation (Eq. \ref{eqmaslope}), and assume another relation between Alfvén Mach number ${\cal M}_{\rm A}$ and density $\rho$ as a power low:
\begin{equation}\label{eqmarho}
    \frac{{\cal M}_{\rm A}}{{\cal M}_{\rm A,c}} = \left(\frac{\rho}{\rho_c}\right)^\gamma \;,
\end{equation}

The fundamental $B-\rho$ relation can be derived by combining fundamental ${\cal M}_{\rm A}$-k$_{B-\rho}$ relation (Eq. \ref{eqmaslope}) with the empirical ${\cal M}_{\rm A}$-$\rho$ relation (Eq.\ref{eqmarho}) :
\begin{equation}\label{eqBrho}\large
    \frac{B}{B_c} = {\rm exp}\left(\left(\frac{\gamma}{{\cal K}}\right)^{-1}\left( \frac{\rho}{\rho_c}\right)^\frac{\gamma}{{\cal K}}\right) \; ,
\end{equation}
and by fitting Eq.\ref{eqBrho} to the observational data from \cite{2010ApJ...725..466C}, assuming ${\cal K}$ =1.7$\pm$0.15 (Sec.\ref{sect2.1}), we find $\gamma = 0.19 \pm 0.07$, ${\rm log_{10}} \rho_{\rm c} /{\rm (g\,cm^{-3})} = 16.09 \pm 1.02$, and ${\rm log_{10}} B_{\rm c}/ {\rm (G)} = -6.29 \pm 0.65$. 
This relation (Eq.\,\ref{eqBrho}) perfectly crosses all Zeeman observations and three MHD simulations and agrees to the classical $B-\rho$ relation \citep{2010ApJ...725..466C,2019ApJ...871...98Z,2020ApJ...890..153J}.
In the diffuse region, the B-field slowly grows with density increasing.
{\en We note that the flat slope at the low-density end can also be caused local variations of the $\cal{M}_A$ at different locations \citep{2024MNRAS.530.3431V,2024arXiv240718293W}.}
In high density region ($\rho\,\sim\,10^{-18}$\,g\,cm$^{-3}$), the slope of $B-\rho$ relation is also up to 2/3, which is caused by star formation \citep{2019ApJ...871...98Z,2021Galax...9...41L}.

Our fitting implies
\begin{equation}
    \frac{{\cal M}_{\rm A}}{{\cal M}_{\rm A,c}} = (\frac{\rho}{\rho_c})^{0.19 \pm 0.07} \;,    
\end{equation}
which is a general relation connecting the degrees of magnetization of clouds in the Milky Way with the density. 
At low density (10$^{-22}$ g\,cm$^{-3}$), the magnetic field is non-negligible. As the density increases, the importance of the magnetic field is gradually diminished. The high-density regions are mostly kinematically-dominated. 
This flat ${\cal M}_{\rm A}-\rho$ relation (${\cal M}_{\rm A}\,\propto$ $\rho^{0.19 \pm 0.07}$) also agrees to actual observations (see the Fig.2 of \citealt{2023ASPC..534..193P}). 
{\en It is similar to the synthesized of slow mode and fast mode in MHD turbulence \citep{2003A&A...398..845P,2024MNRAS.530.3431V}, where the slow mode dominates at low ${\cal M}_{\rm A}$ and low densities, and disappears at high densities. The fast mode dominates at large low ${\cal M}_{\rm A}$ and high densities.}

This gradual decrease in the importance of the magnetic field is consistent with observational data. At the low-density end, the outer region of Taurus \citep{2016MNRAS.461.3918H,2016MNRAS.462.3602T} and Musca \citep{2018Sci...360..635T} molecular clouds have striations (MHD wave, the feature of strong magnetic field state), indicative of strong, dynamically-important magnetic field. At intermediate densities, the increase of the gas density can be accompanied by an increase of the mass-to-flux ratio, and a gradual decrease of the importance of the magnetic field, as observed in the L1544 region \citep{2022Natur.601...49C}, and most high-density regions appears to be kinematically-dominated, such as NGC 6334 \citep{2015Natur.520..518L}, Ser-emb\,8 \citep{2009ApJS..182..143M,2014ApJS..213...13H}, and W51 \citep{2018ApJ...855...39K}.

\section{Conclusion}


We find that the slope of the magnetic field-density relation  k$_{B-\rho}$ = ${\rm d (log_{10} B)}/{\rm d (log_{10} \rho)}$ linked to the  Alfvén Mach number, ${\cal M}_{\rm A}$ using numerical simulation results, 
\begin{equation}
    \rm \frac{{\cal M}_{\rm A}}{{\cal M}_{\rm A,c}} = k_{B-\rho}^{\cal K} \approx \frac{{\cal M}_{\rm A}}{7.5} \approx k_{B-\rho}^{\rm 1.7 \pm 0.15} \nonumber,
\end{equation}
where the ${\cal K}$ = 1.7$\pm$0.15, and the ${\cal M}_{\rm A,c}$ is the characteristic ${\cal M}_{\rm A}$ ($\sim$ 7.5$\pm$1.3). This fundamental relation results from the interplay between turbulence cascade and the implication of magnetic field in MHD turbulence. 
This fundamental relation allows us to estimate ${\cal M}_{\rm A}$ using the slope of the $B-\rho$ relation. We apply this technique to several regions and find that the estimated ${\cal M}_{\rm A}$ values exhibit clear correlations between  ${\cal M}_{\rm A}$ with the observed star formation activities. 

We find that the empirical  $B-\rho$ relation can be explained using our fundamental relation, together with a relation between  ${\cal M}_{\rm A}$ and gas density. 
The ${\cal M}_{\rm A}-\rho$ relation reads
\begin{equation}      
  \rm \frac{{\cal M}_{\rm A}}{{\cal M}_{\rm A,c}} = \left(\frac{\rho}{\rho_c}\right)^\gamma \approx \frac{{\cal M}_{\rm A}}{7.5} \approx \left(\frac{\rho}{10^{-16.1} {\rm g\,cm^{-3}}}\right)^{\rm 0.19\pm 0.07} \; \nonumber,
\end{equation}
and the $B-\rho$ relation takes the following form
\begin{equation}
    \frac{B}{B_c} = {\rm exp}\left(\left(\frac{\gamma}{{\cal K}}\right)^{-1}\left( \frac{\rho}{\rho_c}\right)^\frac{\gamma}{{\cal K}}\right) \approx \frac{B}{10^{-6.3} {\rm G}} \approx {\rm exp}\left(9  \left(\frac{\rho}{10^{-16.1} {\rm g\,cm^{-3}}}\right)^{0.11} \right)\; \nonumber,
\end{equation}
which provides a good fit to existing observational data. 
This fitting result reveals a general trend where a general decrease in the importance of the magnetic field accompanies the increase in the gas density.

The fundamental ${\cal M}_{\rm A}-{\rm k}_{B-\rho}$ relation provides an independent way to estimate the importance of the magnetic field, and it can be widely applied to further observations. The decreasing importance of magnetic field at higher density appears consistent with most of the existing observations and will be tested against future results.

\section*{Acknowledgements}

We thank the referee for the careful reading of the paper and the constructive comments. 
We thank Enrique V{\'a}zquez-Semadeni for the helpful comments.
GXL acknowledges support from the National Natural Science Foundation of China (NSFC) grants No. 12273032 and 12033005. 
K.Q. acknowledges support from NSFC grant No.12425304, U1731237, and National Key R$\&$D Program of China 2023YFA1608204 and No. 2022YFA1603100.

\bibliography{reference}

\appendix



\newpage

\textbf{\Large{Method}}

\section{Simulation}\label{enzo}

The numerical simulation of molecular cloud selected in this work is identified within the three $\beta_0$ simulations applied by the constrained transport MHD option in Enzo (MHDCT) code \citep{2010ApJS..186..308C,2015ApJ...808...48B}. 
The simulation conducted in this study analyzed the impact of self-gravity and magnetic fields on supersonic turbulence in isothermal molecular clouds, using high-resolution simulations and adaptive mesh refinement techniques. which detail in shown in \citep{2012ApJ...750...13C,2015ApJ...808...48B,2020ApJ...905...14B}. 
The Enzo simulation with three dimensionless parameters provides us with massive information on physical processes:
\begin{equation}
    {\cal M}_s = \frac{v_{\rm rms}}{c_s} = 9
\end{equation}
\begin{equation}
    \alpha_{vir} = \frac{5 v_{\rm rms}^2 }{ 3 G \rho_0 L_0^2} = 1
\end{equation}
\begin{equation}\label{eq3}
    \beta_0 = \frac{8\pi c_s^2 \rho_0}{B_0^2} = 0.2, 2, 20
\end{equation}
where $v_{\rm rms}$ is the rms velocity fluctuation, c$_s$ is sonic speed, $\rho_0$ is mean density, and $B_0$ is the mean magnetic field strength.
With the same sonic Mach number ${\cal M}_{\rm s}$ and viral parameter $\alpha_{vir}$ in initial conditions, the difference only exists in initial magnetic pressure $\beta-0$ with various B-field strength,
The $\beta_0$ as 0.2, 2, and 20 present the strong B-field state, medium B-field, and weak B-field state.
The clouds are in three different evolutionary stages started with self-gravity existing at 0.6$t_{\rm ff}$($\sim$ 0.76\,Myr), where $t_{\rm ff}$ represents 1.1$((n_{H} / 10^3)^{-1/2}$\,Myr \citep{2012ApJ...750...13C,2015ApJ...808...48B}. 
Due to the short timescale of evolution, the gravitational collapse could barely affect the B-$\rho$ relation in the simulation
The size of these molecular clouds is around 4.6$^3$ pc (256$^3$ pixels), with one-pixel size of approximately 0.018 pc \citep{2015ApJ...808...48B}.
Fig.\,\ref{figBrho} show the distribution of B-field and density with three $\beta_0$ in the timescale of 0.6 $t_{\rm ff}$ ($\sim$ 0.76 Myr).


In this Enzo simulation, the thermal energy is located at the low state (${\cal M}_s$ = 9).
Their timescale of evolution stage is 0.6 $t_{\rm ff}$, around  0.76 Myr, which the timescale starts from when interstellar media collapse.
Due to the short evolutionary time ($\sim$ 3.85 $\times$ 10$^{-21}$ g\,cm$^{-3}$, or 821 cm$^{-3}$), the gravitational energy less affects the system in simulation.
With the similar mean density $\rho_0$ in these simulations, the different $\beta_0$ ($\sim$0.2, 2, 20) present the different magnetic field states, strong B-field, stronger B-field (weaker than strong B-field but stronger than weak B-field), and weak B-field.
The energy ratio between kinetic energy $E_{\rm k}$ and magnetic energy $E_{\rm B}$ $E_{\rm k}$/$E_{B}$ in the evolutionary timescale of 0.6\,t$_{ff}$ ($\sim$ 0.76 Myr) is calculated:
\begin{equation}
    \frac{E_{\rm k}}{E_{\rm B}} = \frac{\frac{1}{2}\rho v_{\rm rms}^2}{B_0^2/8\pi} 
\end{equation}
The $E_{\rm k}$/$E_{B}$ with stronger, medium, and weak B-field states are 0.09, 0.24, and 1.66, respectively.

\section{Re-scaling method}\label{aprescalling}
The values obtained from numerical simulations are without dimensions, and the numerical values of the physical quantities are obtained during the interpretation phase, where the pre-factors are inserted to ensure the dimensionless parameters such as the Mach number stay unchanged. When applying the result of numerical simulations to reality, one is allowed to change the numerical values of the prefactors, as long as the dimensionless controlling parameters stay unchanged. 

Each simulation has three dimensionless controlling parameters, including the Mach number, the plasma $\beta_0$, and the virial parameter. We argue that the virial parameter has little influence on the slope of the magnetic-field-density relation. This is because gravity is not important in determining the relation. This assertion is supported by the fact that the magnetic field-density relation do not evolve significantly during the collapse phase.


We thus rescale our simulations using the following transform:
\begin{equation}
    B' = \lambda_B \cdot B,\,\,\,\,\ \rho' = \lambda_\rho \cdot \rho
\end{equation}
where $\lambda_B$ and $\lambda_\rho$ are the rescalling factors, the B and $\rho$ are the B-field and density before rescalling, the B' and $\rho'$ are the B-field and density after rescalling. 
In this simulation, initial ${\cal M}_{\rm A,0}$ is amount to the plasma $\beta_0$, ${\cal M}_{\rm A_0} = (\rho_0 v_{\rm rms}^2/B_0^2)^{0.5} = (81 c_s^2 \rho_0 /B_0^2)^{0.5} = (\frac{81}{8\pi}\beta_0)^{0.5}$.
With evolving, the unchanged of ${\cal M}_{\rm A}$ means that plasma $\beta_0$ is not changed.
We thus require that this rescaling ensure that the Alfvenic Mach number ${\cal M}_{\rm A}$ of simulations do not change.


The density after rescalling, $\rho'$, is obey on the ${\cal M}_{\rm A}-\rho$ relation (See Eq.\,\ref{eqmarho}):
\begin{equation}
    \rho' = (\frac{{\cal M}_{\rm A}}{{\cal M}_{\rm A,0}})^{1/\gamma} \cdot \rho_c
\end{equation}
where the $\gamma$ is the slope of ${\cal M}_{\rm A}$-$\rho$ empirical relation, the ${\cal M}_{\rm A,c}$ is the characteristic ${\cal M}_{\rm A}$ (see Eq.\,\ref{eqmarho}).
The rescalling factor of density is calculated as:
\begin{equation}
    \lambda_\rho = \frac{\rho'}{\rho} = (\frac{{\cal M}_{\rm A}}{{\cal M}_{\rm A,0}})^{1/\gamma} \cdot \frac{\rho_c}{\rho}
\end{equation}
where the $\rho$ is density before rescalling.
The B-field strength after rescalling, B', obeys the $B-\rho$ relation (See Eq.\,\ref{eqBrho}) and is calculated by rescalling density $\rho'$:
\begin{equation}
    B' = {\rm exp}((\frac{\gamma}{{\cal K}})^{-1}( \frac{\rho'}{\rho_c})^\frac{\gamma}{{\cal K}}) \cdot B_c
\end{equation}
where the parameter, $B_c$, $\rho_c$, $\gamma$, and ${cal K}$, are the same as Eq.\,\ref{eqBrho}, the B is the B-field before scalling.
The B-field rescalling factor $\lambda_B$ is defined as:
\begin{equation}
    \lambda_B = \frac{B'}{B} = {\rm exp}((\frac{\gamma}{{\cal K}})^{-1}( \frac{\rho'}{\rho_c})^\frac{\gamma}{{\cal K}}) \cdot \frac{B_c}{B}
\end{equation}
where the $B$ is B-field before rescalling.
The parameters before or after rescalling are shown in Tab.\,\ref{tabpbfr}.
The rescalling factors are shown in tab.\,\ref{tablamda}.

{\en After re-scaling, the $B-\rho$ distributions in the simulation with different initial $\beta_0$ form an overall trend (see Fig.\,\ref{figBrho}), similar to the slow mode at low density and the fast mode at high density \citep{2003A&A...398..845P,2024MNRAS.530.3431V}.
The ${\cal M}_{\rm A}$ $\approx$ 1 is close to the transition point in t
he classical $B_\rho$ relation \citep{2010ApJ...725..466C}, which could be a consequence of the diversity of parameters in the different database entries, such as ${\cal M}_{\rm A}$ and ${\cal M}_{\rm s}$ \citep{2003A&A...398..845P}, and the superposition of the slow and fast mode scalings \citep{2024arXiv240718293W,2024MNRAS.530.3431V}.}

\begin{table}[]
    \centering
    \caption{The physical Parameters of simulation before and after rescalling.}
    \begin{tabular}{|c|c c c|c c c|}
        \hline
         &   \multicolumn{3}{c|}{Before Rescalling} &   \multicolumn{3}{c|}{After Rescalling}\\
        simulation &  $\beta_0$ = 0.2 & $\beta_0$ = 2 & $\beta_0$ = 20 & $\beta_0$ = 0.2 & $\beta_0$ = 2 & $\beta_0$ = 20 \\
        \hline
        ${\cal M}_{\rm A}$ & 0.31 & 0.49 & 1.29 & 0.31 & 0.49 & 1.29 \\
        $\Bar{\rho}$ (g cm$^{-3}$) & 3.9$\times$10$^{-21}$ & 3.9$\times$10$^{-21}$ & 3.9$\times$10$^{-21}$ & 4.3$\times$10$^{-24}$ & 4.8$\times$10$^{-23}$ & 7.8$\times$10$^{-21}$ \\
        $\Bar{B}$ (G) & 2.8$\times$10$^{-5}$ & 1.6$\times$10$^{-5}$ & 8.4$\times$10$^{-6}$ & 2.2$\times$10$^{-6}$ & 3.5$\times$10$^{-6}$ & 1.4$\times$10$^{-5}$ \\
        $E_{\rm k}$ / $E_{\rm B}$ & 0.09 & 0.24 & 1.66 & 0.09 & 0.24 & 1.66 \\
        \hline
    \end{tabular}
    \label{tabpbfr}
\end{table}

\begin{table}[]
    \centering
    \caption{The rescalling factors $\lambda_{B}$ and $\lambda_{rho}$.}
    \large
    \begin{tabular}{|c|c c|}
         \hline
         simulations & log$_{10}$ $\lambda_{B}$ & log$_{10}$ $\lambda_{rho}$ \\
         \hline
         $\beta_0$ = 0.2 & -1.1 & -2.96  \\
         $\beta_0$ = 2 & -0.67 & -1.91  \\
         $\beta_0$ = 20 & 0.24 & 0.3  \\
         \hline
    \end{tabular}
    \label{tablamda}
\end{table}

\section{Fitting slope of $B-\rho$ Distribution}

\begin{figure}
    \centering
    \includegraphics[width = 13cm]{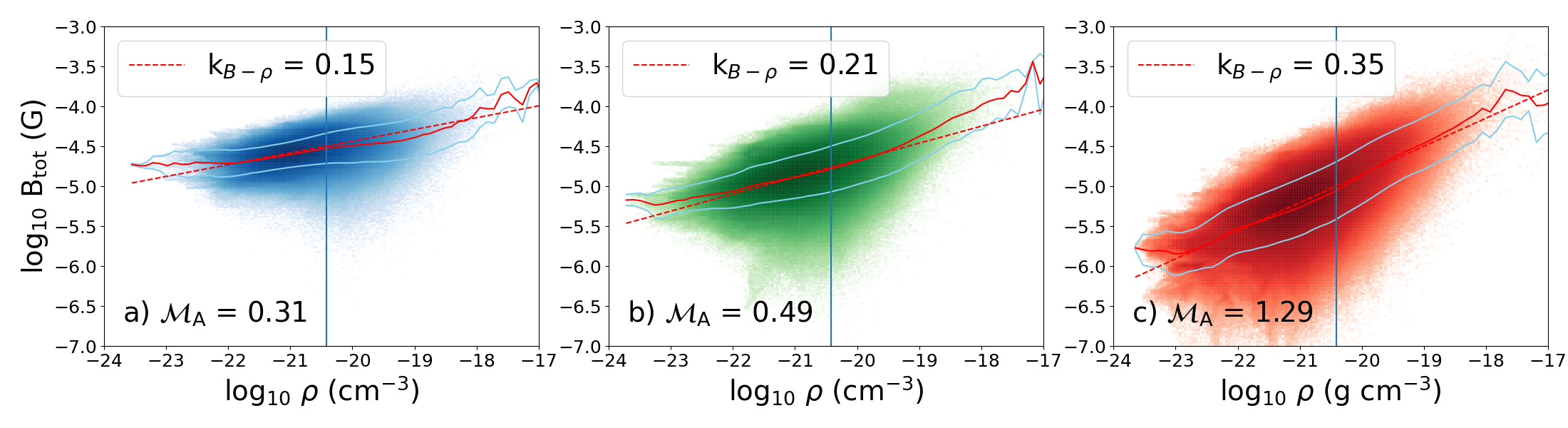}
    \caption{\textbf{Fitting the slope of B-$\rho$ relation in log-log space.}
    The a, b, and c panels display the B-$\rho$ distribution of simulations with $\beta_0$ = 0.2, 2, and 20, respectively.
    The red solid lines show the average B-$\rho$ relation and the blue lines show the dispersion region of the average B-$\rho$ relation.
    The red dashed lines are linear equations fitting the average B-$\rho$ relation.
    The blue perpendicular lines present the mean density $\Bar{\rho}$.
    The slope of the $B-\rho$ relation is fitted around the mean density of the simulation as 0.15, 0.21, and 0.35, respectively.}
    \label{figslopesims}
\end{figure}

Fig.\,\ref{figslopesims} show the distribution between B-field and density in simulation with ${\cal M}_{\rm A}$ = 0.31, 0.49, 1.29, respectively.
Fitting the main structure of $B-\rho$ distribution at the main density range of simulation in log-log space, the slope of $B-\rho$ relation, ${\rm d (log_{10} B)}/{\rm d (log_{10} \rho)}$ in simulations (${\cal M}_{\rm A}$ = 0.31, 0.49, 1.29) can be derived as 0.15, 0.21, 0.35, respectively. 

table, with all  parameters, energy ratios,



\section{Calculated the Slope of $B-\rho$ relation in Observation}\label{apslope}

\begin{figure}
    \centering
    \includegraphics[height = 9cm]{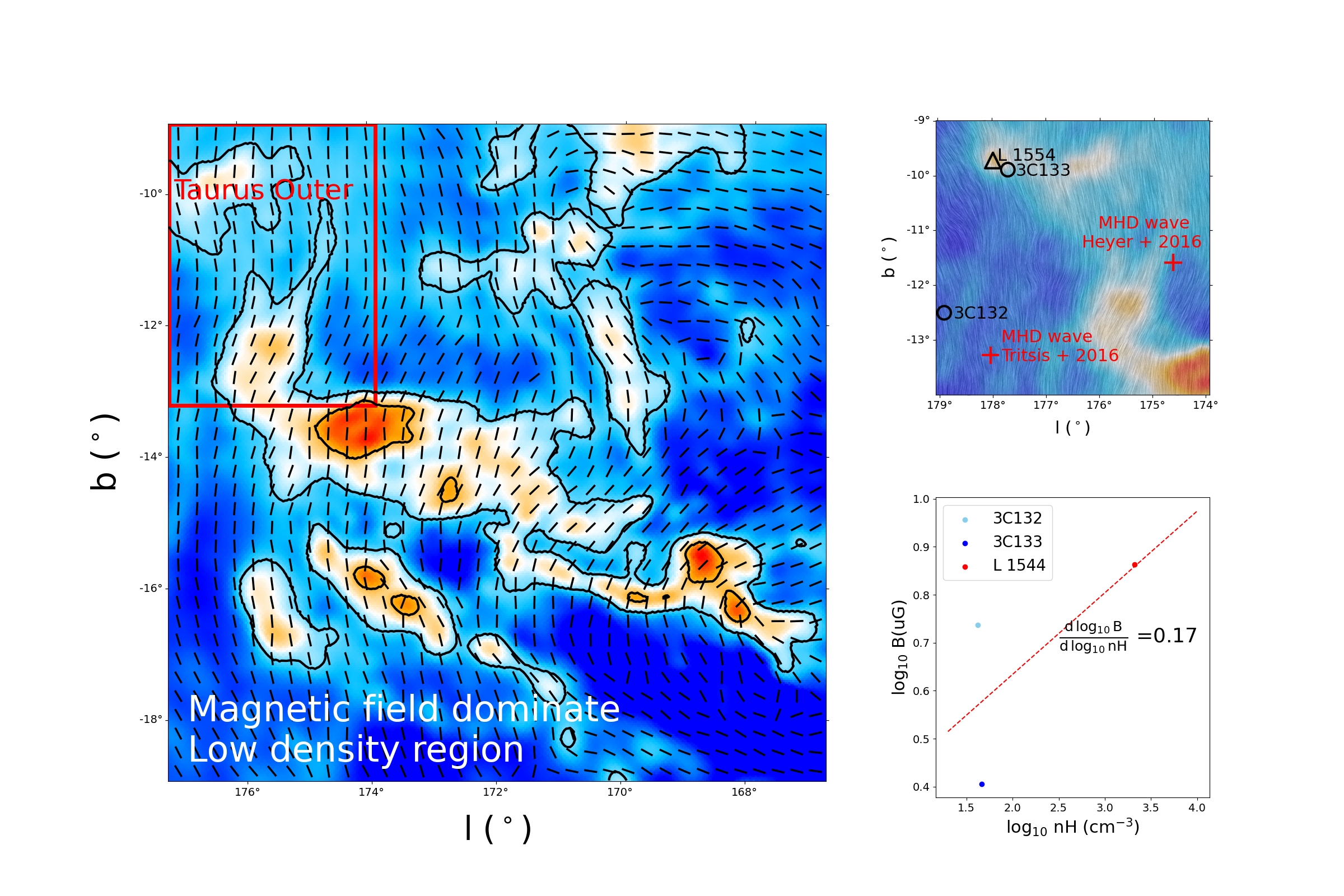}
    \caption{\textbf{Fitting $B-\rho$ slope of Taurus outer region.}
    The left panel displays the magnetic field of Taurus derived by Planck 353 GHz dust polarization \citep{2020A&A...641A..12P}.
    The contours show the region where column density is above 2$\times$ 10$^{21}$ cm$^{-2}$. 
    Other region, column density $\textless$ 2$\times$ 10$^{21}$ cm$^{-2}$, could exist the MHD wave \citep{2016MNRAS.461.3918H}.
    The red region shows the Taurus outer region we used in this work.
    The right top panel shows the magnetic field morphology of the Taurus outer region and the position of distribution of 3C132, 3C133, and L1544.
    The right bottom panel displays the sources measured with the Zeeman effect located at the Taurus outer region, 3C132, 3C133, and L1544 in B-nH space, which show the mean magnetic field B and mean H$_{2}$ volume density nH \citep{2010ApJ...725..466C,2022Natur.601...49C}.
    The right panel shows the positions of sources measured with Zeeman effect, 3C132, 3C133, TMC-1, and L1544 in l-b space \citep{2004ApJS..151..271H,2008ApJ...680..457T}.
    The crosses show the candidate of MHD wave sources in the Taurus outer region \citep{2016MNRAS.461.3918H,2016MNRAS.462.3602T}.}
    \label{figTaurus}
\end{figure}

Calculating Slope of $B-\rho$ relation have two method: \\
i) Single source with various scales \\
ii) Single source with various densities \\

\subsection{Single source with various scales}

The B-field $B$ and density $\rho$ of a single source is observed at various scales, such as NGC\,6334 \citep{2015Natur.520..518L}(${\rm d (log_{10} B)}/{\rm d (log_{10} \rho)}$ = 0.41).

\subsection{Single source with various densities}

The region of a single source has sub-regions with various densities. 
For example, the diffuse region (N(H$_2$) $\textless$ 2$\times$ 10$^{22}$ cm$^{-2}$) in the Taurus could exist MHD wave \citep{2016MNRAS.461.3918H}.
As Fig.\,\ref{figTaurus} shows, the Taurus out region has sub-areas with various densities, 3C132, 3C133, L1544 \citep{2016MNRAS.461.3918H,2016MNRAS.462.3602T}.
Zeeman source 3C132 is located in the MHD Wave region and 3C133, L1544 is located at the boundary line of the MHD Wave region.
\citep{2022Natur.601...49C} found that L1544 exists as the transition point between strong and weak magnetic fields.
Fig.\,\ref{figTaurus} shows the distribution between the mean density and mean B-field of three sources, which mean $\rho$ and $B$ come from \citep{2010ApJS..186..308C,2022Natur.601...49C}. 
The slope of the $B-\rho$ relation in the Taurus outer region is fitting as 0.17. 

\section{Energy Spectrum $\And$ Velocity Dispersion}

\begin{figure}
    \centering
    \includegraphics[width = 10cm]{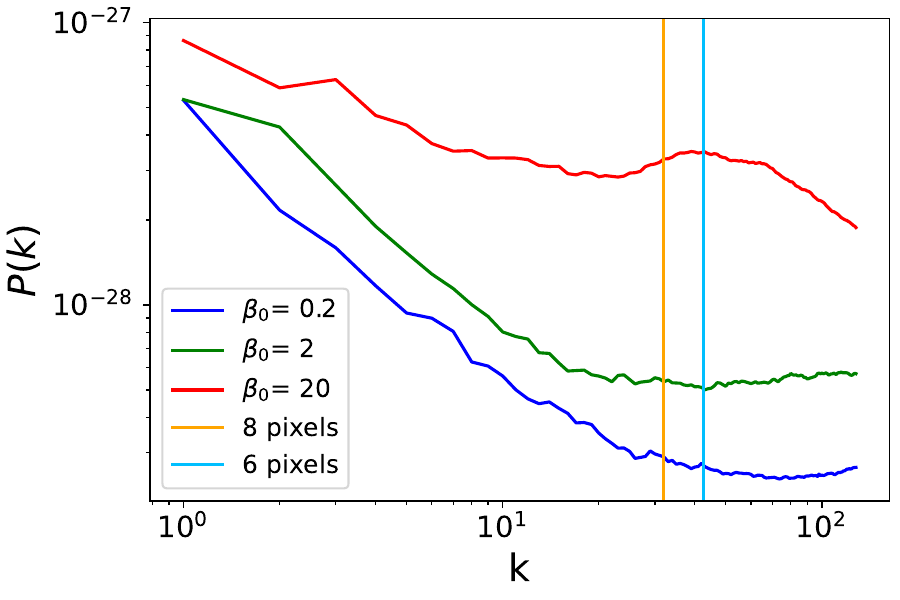}
    \caption{{\bf Energy spectrum of density in situations with various $\beta_0$}.
    The blue, green, and red lines show the energy spectrum of density with $\beta_0$ as 0.2, 2, and 20, respectively.
    The orange and vertical blue lines display the structure on the scale of 8, and 20, respectively.}
    \label{fig6}
\end{figure}

The energy spectrum of density $\rho$ in three simulations presents the density structure focused on the 6-pixel scale (see Fig.\,\ref{fig6}). 
We use the 8-pixel as the sub-block size to calculate the velocity dispersion $\sigma_v$ in each pixel of simulation data, which is close to the 6-pixel scale and can divide the whole scale of simulation data (256 pixels) without remainder:
\begin{equation}
    v_{ijk, m} = \frac{\sum v_{ijk} \cdot \rho}{\sum \rho} \sum v_{ijk} 
\end{equation}
\begin{equation}
    \sigma_{v} = (\sum (v_{ijk} - v_{ijk, m})^2)^{\frac{1}{2}}
\end{equation}
where the i,j,k show the axis of x,y, and z.
The velocity dispersion can be used to compute the kinetic energy density ($e_k = \frac{1}{2}\rho\sigma_v^2$).

\end{document}